\documentclass[twocolumn,aps,prd]{revtex4-1}
\usepackage{graphicx}
\usepackage{bm}
\usepackage{relsize}
\usepackage{amsmath,amsfonts,amsthm}

\begin{document}
\newcommand*{\bi}{\bibitem}
\newcommand*{\ea}{\textit{et al.}}
\newcommand*{\eg}{\textit{e.g.}}
\newcommand*{\zpc}[3]{Z.~Phys.~C \textbf{#1}, #2 (#3)}
\newcommand*{\plb}[3]{Phys.~Lett.~B \textbf{#1}, #2 (#3)}
\newcommand*{\phrc}[3]{Phys.~Rev.~C~\textbf{#1}, #2 (#3)}
\newcommand*{\phrd}[3]{Phys.~Rev.~D~\textbf{#1}, #2 (#3)}
\newcommand*{\phrl}[3]{Phys.~Rev.~Lett.~\textbf{#1}, #2 (#3)}
\newcommand*{\pr}[3]{Phys.~Rev.~\textbf{#1}, #2 (#3)}      
\newcommand*{\npa}[3]{Nucl.~Phys.~A \textbf{#1}, #2 (#3)}  
\newcommand*{\npb}[3]{Nucl.~Phys.~B \textbf{#1}, #2 (#3)}  
\newcommand*{\npbps}[3]{Nucl.~Phys.~B (Proc. Suppl.) \textbf{#1}, #2 (#3)}  
\newcommand*{\ptp}[3]{Prog. Theor. Phys. \textbf{#1}, #2 (#3)}
\newcommand*{\ppnp}[3]{Prog. Part. Nucl. Phys. \textbf{#1}, #2 (#3)}
\newcommand*{\ibid}[3]{\textit{ ibid.} \textbf{#1}, #2 (#3)}
\newcommand*{\epjc}[3]{Eur. Phys. J. C \textbf{#1}, #2 (#3)}
\newcommand*{\jpg}[3]{J. Phys. G \textbf{#1}, #2 (#3)}
\newcommand*{\mpla}[3]{Modern Physics Letters A \textbf{#1}, #2 (#3)}
\newcommand*{\ra}{\rightarrow}
\newcommand*{\pippim}{\pi^+\pi^-}
\newcommand*{\kpkm}{K^+K^-}
\newcommand*{\kskl}{K^0_SK^0_L}
\newcommand*{\rf}[1]{(\ref{#1})}
\newcommand*{\be}{\begin{equation}}
\newcommand*{\ee}{\end{equation}}
\newcommand*{\bea}{\begin{eqnarray}}
\newcommand*{\eea}{\end{eqnarray}}
\newcommand*{\nl}{\nonumber \\}
\newcommand*{\rmd}{\mathrm d}
\newcommand*{\die}{e^+e^-}
\newcommand*{\jj}{\mathrm i}
\newcommand*{\ndf}{\mathrm{NDF}}
\newcommand*{\mev}{\mathrm{~MeV}}
\newcommand*{\cndf}{\chi^2/\mathrm{NDF}}
\newcommand*{\minuit}{\texttt{MINUIT}~}
\newcommand*{\w}{\sqrt s}
\newcommand*{\e}[1]{{\mathrm e}^{#1}}
\newcommand*{\ie}{\textrm{i.e.}}
\newcommand*{\dek}[1]{\times10^{#1}}
\newcommand*{\chreact}{$e^+e^- \rightarrow K^+ K^-$}
\newcommand*{\nreact}{$e^+e^- \rightarrow K_S K_L$}
\def\babar{\mbox{\slshape B\kern-0.1em{\smaller A}\kern-0.1em
    B\kern-0.1em{\smaller A\kern-0.2em R}}}

\title{Manifestation of kaonium in the 
$\bm{e^+e^- \rightarrow K^+ K^-}$ process}
\author{Peter Lichard}
\affiliation{
Institute of Physics and Research Centre for Computational Physics
and Data Processing, Silesian University in Opava, 746 01 Opava, 
Czech Republic\\
and\\
Institute of Experimental and Applied Physics,
Czech Technical University in Prague, 128 00 Prague, Czech Republic
}
\begin{abstract}
We analyze the precise data obtained by the CMD-3 experiment
on the $e^+e^-$ annihilation into two charged kaons in the 
vicinity of the $\phi$ peak. A perfect fit is obtained only if a pole
on the real axis below the reaction threshold is assumed. This can be
interpreted as proof of the existence of the 2p state of kaonium, a compound
of $K^+$ and $K^-$. The \babar~Collaboration data on the same process supports 
this conclusion and, in addition, points to the strong interaction as
a dominant source of the binding energy.
The possibility of discovering 
2p kaonium in the $e^+e^-\rightarrow \pi^+\pi^-$ process is discussed.
The 2p state of $K^0$-onium is indicated on the basis 
of the CMD-3 $e^+e^-\rightarrow K^0_SK^0_L$ data.
\end{abstract}
\maketitle
\section{Introduction}
Kaonium is still a hypothetical compound system consisting of a positively
charged and a negatively charged kaon. It belongs to a wide class of onia,
systems made of a particle and an antiparticle. In the lepton sector,
they are the well-known positronium, true muonium, and true tauonium 
(the last two have yet to be observed). Quarkonia, the bound states of a 
quark and its antiquark, are observed as truly neutral (all flavor quantum
numbers vanishing) mesons. They are numerous and include, \eg,
$\phi$ ($s\bar s$), $\eta_c$ and $J/\psi$ ($c\bar c$), $\Upsilon$ ($b\bar
b$). 
Many theoretical studies, starting with the Fermi-Yang and Sakata 
models \cite{fys}, have considered the possibilities of baryon-antibaryon 
bound states.

In the meson sector, pionium ($\pi^+\pi^-$) was discovered in 1993 at
the 70~GeV proton synchrotron at Serpukhov, Russia \cite{pionium}, and 
intensively studied in the Dimeson Relativistic Atomic Complex (DIRAC) 
experiment~\cite{shortlife} at the CERN Proton Synchrotron. Assuming pure 
Coulombic interaction, the binding energy of pionium can be calculated from 
the hydrogen-atom formula 
\be
\label{hatom}
b_n=\frac{m_r\alpha^2}{2n^2},
\ee
where $m_r$ is the reduced mass in energy units (used throughout this paper), 
$\alpha\approx 1/137$ is the fine-structure constant, and $n$ is the principal 
quantum number. Putting $n=1$ for the ground state and $m_r=m_{\pi^+}/2$, we 
get $b=1.86$~keV. The decay to two neutral pions is dominant and 
the measured lifetime is $3.15^{+0.28}_{-0.26}\dek{-15}$~s 
\cite{shortlife}. The NA48/2 Collaboration at the CERN Super Proton
Synchrotron~\cite{NA48/2} studied decays $K^\pm\ra\pi^\pm\pi^0\pi^0$ and 
found an anomaly in the $\pi^0\pi^0$ invariant mass distributions that can 
be interpreted as the production of pionia in the kaon decays and their 
subsequent two-$\pi^0$ decay. 

The DIRAC experiment also observed  and studied $\pi^-K^+$ and  $\pi^+K^-$ 
atoms~\cite{pikevid,pikobser}.

To date, the experiments concerning dimeson production have been performed at
proton accelerators
\cite{pionium,shortlife,NA48/2,pikevid,pikobser,longlife}. 
Electron-positron colliders, the machines that are famous for participating 
in the discovery of many new particles (notably quarkonia), have not yet 
contributed much to mesonia physics. The reason is that ground-state mesonia 
(1s in atomic notation) are objects with $J^{PC}=0^{++}$ quantum numbers, and 
as such they cannot couple to the photon.
In the $\die$ annihilation processes, they must be accompanied by an additional
particle or particles, or at least a photon \cite{notescalar}.
However, the DIRAC Collaboration recently discovered \cite{longlife} so-called
long-lived $\pippim$ atoms, which are the 2p atomic states with quantum
numbers $J^{PC}=1^{--}$. Therefore, they can be produced in the $\die$ 
collisions. The Coulombic binding energy of the 2p pionium is 0.464~keV;
its lifetime was determined in Ref.~\cite{longlife} to be 
$\tau_{2p}=0.45^{+1.08}_{-0.30}\dek{-11}$~s. Such a long lifetime is caused 
by the fact that the decay modes to the positive C-parity states $\pi^0\pi^0$ 
and $\gamma\gamma$ are now forbidden and the 2p$\to$1s transition
dominates \cite{pionium2pto1s}.

No experimental evidence of kaonium has yet been found. In the simplest way, 
kaonium can be considered a hydrogenlike atom (a system held together due to 
Coulombic attraction between opposite electric charges). Equation \rf{hatom} 
gives the binding energy of the ground state $b\approx 6.57$~keV. Unlike the 
hydrogen atom and leptonic onia, the constituents of pionium and kaonium 
also interact via strong force. Krewald \ea~\cite{krewald2004} found 
``the ground state energy for the kaonium atom that is shifted above the 
Coulomb value by a few hundred eV." As a rule, an increase in bound-state 
energy means a drop in binding energy. On the contrary, Zhang 
\ea~\cite{zhang2006} found that kaonium binds more strongly ($b=7.05$~keV) 
than it corresponds to Coulomb interaction. 

Kaonium is not stable. Ground-state (1s) kaonium partly decays 
electromagnetically into two photons. In addition, the exchange of $K^*$ 
between kaons generates the $\pi^+\pi^-$, $\pi^0\pi^0$, and $\eta\pi^0$ decay 
modes. Klevansky and Lemmer~\cite{klevlemm2011} used meson-meson interaction 
amplitudes taken from leading order chiral perturbation theory and found the 
resulting lifetime of $(2.2\pm0.9)\dek{-18}$~s. This is in conformity with
the order of magnitude estimate $10^{-3}$~fs obtained by Deloff~\cite{deloff}.

The Coulombic binding energy of the first excited state ($n=2$) of kaonium
is 1.64~keV. We will concentrate on the 2p state, which can be produced in the
$\die$ experiments. Its quantum numbers $J^{PC}=1^{--}$ forbid decays to
the $C$=1 $\pi^0\pi^0$ and $\eta\pi^0$ states. The 2p kaonium width
is thus determined by the decay rate into $\pi^+\pi^-$ 
only~\cite{notegammatransition}. Its lifetime should be at least 2 times higher
than that of the ground-state kaonium. The corresponding decay width  
is around 0.1~keV, about 4 orders of magnitude smaller than 
the decay width of the $\phi(1020)$, which is the dominant object in the 
process we are going to investigate. For our purposes, we can thus neglect 
the 2p kaonium decay width and consider it a stable particle. 

\section{2$\bm{p}$ kaonium as a pole in the $\bm{\die\to\kpkm}$ amplitude}
In this paper, we analyze the existing precise data on the $\die\ra\kpkm$ 
process with the aim of finding a pole in the amplitude corresponding to a
bound state. 

The reaction amplitude is a function of $s$, the invariant energy squared. 
Under very general conditions, the amplitude can be continued into the 
complex $s$-plane. The resonances are represented by the poles at imaginary 
$s$. The (quasi) stable particle, or bound state,  appears as a pole at real 
$s$ below the reaction threshold.

The formula for the cross section of the $\die$ annihilation into a $K^+K^-$ 
pair based on the vector-meson-dominance model with two resonances is  
\bea
\label{sigma}
\sigma(s)&=&\frac{\pi\alpha^2}{3s}\left(1-\frac{4m_K^2}{s}\right)^{3/2}
\nl
&\times&\left|
\frac{R_1e^{\jj\delta}}{s-M_1^2-\jj M_1\Gamma_1}+
\frac{R_2}             {s-M_2^2-\jj M_2\Gamma_2}
\right|^2,
\eea
where $M_i$ and $\Gamma_i$ determine the position and width of the $i$th
resonance, respectively. The residuum $R_i$ includes the product of two
constants. One characterizes the coupling of the $i$th resonance to the 
photon (up to the elementary charge $e$, which is taken off to form, 
after squaring, an $\alpha$ in the prefactor) and the other is the coupling 
of the resonance to the $K^+K^-$ pair. The phase $\delta$ regulates the 
interference between the resonances. 

\begin{figure}[h]
\includegraphics[width=8.6cm]{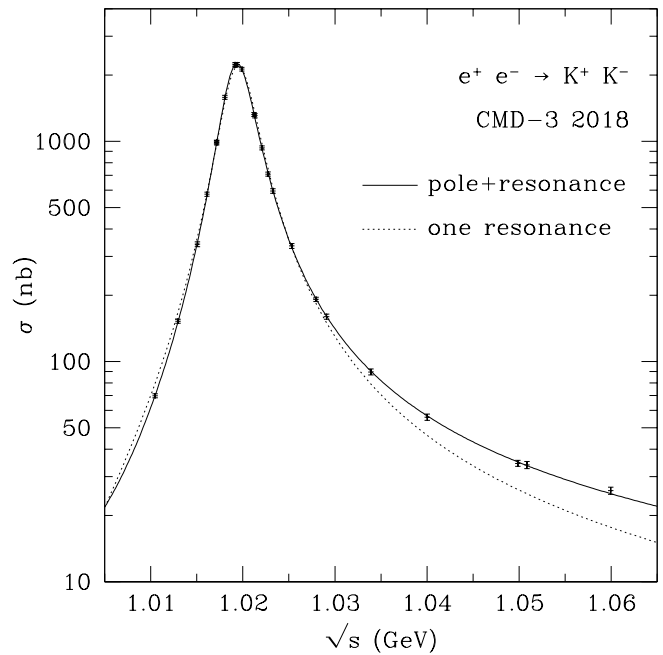}
\caption{\label{fig:kaonium}Cross section for the $\die$ annihilation
into two charged kaons measured in the CMD-3 experiment~\cite{cmd2018}
and two fits to it using Eq. \rf{sigma}. Their parameters are given in
Table~\ref{tab:parameters}.}
\end{figure}

We use the $\die\ra K^+K^-$ cross section data \cite{cmd2018} obtained
by the CMD-3 (Cryogenic Magnetic Detector) experiment at the VEPP-2000 
$\die$ collider in Novosibirsk, Russia. The advantage of the energy scan 
method used in this experiment is in getting high statistics data at precisely
known energies, to which the collider is tuned up step by step. The number 
of data points is 24, and they are concentrated in a narrow
region around the $\phi(1020)$ resonance. The data are shown in 
Fig.~\ref{fig:kaonium}, together with two curves depicting our 
fits~\cite{notecmd2018}.

We first tried to fit the data assuming just one resonance, \ie, using 
three free parameters~\cite{fred}. The result was disastrous: the usual 
$\chi^2$ was equal to 341.8, which together with the number of degrees of 
freedom $\ndf=24-3=21$ implied a confidence level (C.L.) of zero. This fit 
is depicted by a dashed curve in Fig.~\ref{fig:kaonium}. The  parameters of 
the fit are listed in the middle column of Table~\ref{tab:parameters}.
\begin{table}[b]
\caption{\label{tab:parameters}Parameters of the two fits to the CMD-3 
$K^+K^-$ data \cite{cmd2018} depicted in Fig. \ref{fig:kaonium}.}
\begin{tabular*}{8.6cm}[b]{lcc}
\hline
                &~~~One resonance~~~&~~~Resonance and pole~~~\\
\hline
$R_1$ (GeV$^2$) & 0.3679(15)    & 0.3759(15)   \\
$M_1$ (GeV)     & 1.019393(14)  & 1.019247(18) \\
$\Gamma_1$ (MeV)& 4.359(32)     & 4.172(38)    \\
$\delta$        & 0 (fixed)     & 1.334(34)    \\
$R_2$ (GeV$^2$) & 0 (fixed)     & 0.298(23)    \\
$b$ (keV)       &       & 1.64\footnote{$M_2=2m_{K^+}-b$} (fixed)\\
$\Gamma_2$      &               & 0 (fixed)           \\
$\chi^2$        & 341.8         & 4.6          \\
NDF             & 21            & 19           \\
Confidence level     & 0             & 100\%        \\
\hline
\end{tabular*}

\end{table}
It must be said that the Kozyrev \ea~\cite{cmd2018} achieved a better 
result. They used a more sophisticated parametrization of the 
$\phi(1020)$-resonance shape based on an energy dependent width $\Gamma(s)$ 
and attained $\cndf=25/20$, which gave C.L.~$=20$\%.

We then tried to improve our fit by considering two resonances. To our surprise,
the width of the second resonance came out close to zero ($0.6\pm1.7$~MeV
\cite{noteerrorg2}) and the resonance
position was below the threshold, which signalized the bound state. 
Inspired by that, we replaced the second resonance with a bound-state pole
by putting $\Gamma_2\equiv 0$ and $M_2=2m_{K^+}-b$, where
$b$ is the 2p kaonium binding energy. We took $b=1.64$~keV as a first try.
We attained  perfect agreement with the data ($\cndf=4.6/19$,  
C.L.~=~100\%), depicted by the full curve in Fig.~\ref{fig:kaonium}.
All parameters of the fit are listed in the rightmost column of 
Table~\ref{tab:parameters}. Minimizing the $\chi^2$ with respect to the binding 
energy gives an estimate of $b=(10.8^{+14.8}_{-9.1})$~MeV \cite{chisq}. 
Let us mention that in 1961 Uretsky and Palfrey~\cite{uretsky} assumed 
a mesonium binding energy of 10~MeV.

A binding energy much larger that its Coulombic value of 1.64~keV
would mean that the strong force between $K^+$ and $K^-$ is attractive
and responsible for keeping kaonium together. But our result is still not 
conclusive. The parameters of the fit with $b=10.8$~MeV 
($\cndf=2.4/19$, C.L.~=~100\%) are only marginally better than those
with the Coulombic binding energy shown in Table~\ref{tab:parameters}. 
To get more insight into this problem, we explored two other datasets.

The $\die\ra\kpkm$ data obtained by the CMD-2 experiment at the VEPP-2000 
$\die$ collider in Novosibirsk, Russia,  were published in 2008 \cite{cmd2}. 
They contain 21 points in an energy range narrower than that of the later 
CMD-3 data~\cite{cmd2018} explored above. We have performed the fits to 
data~\cite{cmd2} under three different assumptions; the results are shown 
in the middle panel of Table \ref{tab:twoexperiments}. The one-resonance fit 
is already perfect. Adding kaonium with either Coulombic binding energy or 
the Uretsky-Palfrey value of 10~MeV further decreases $\chi^2$. Owing to 
perfect fit without kaonium, this cannot be considered an unambiguous proof 
of kaonium's existence. 
\begin{table}[b]
\caption{\label{tab:twoexperiments}Evidence for 2p kaonium from two other
experiments. The last two rows show results of the fits ($\cndf$,
C.L.) with kaonium considered in addition to the resonance.}
\begin{tabular*}{8.6cm}{lcc}
\hline
                & CMD-2 \cite{cmd2}& \babar~\cite{babar2013}\\
\hline
Energy range (GeV) &~~1.01136--1.03406~~& 0.985--1.065   \\
No. of points    & 21  & 48 \\
One-resonance fit   & 7.5/18~~~~98.5\%  &244.4/45~~~~~0\%\\ 
Kaonium $b=1.64$~keV & 2.2/16~~~~100\%  &184.5/43~~~~~0\% \\      
Kaonium $b=10$~MeV   & 2.3/16~~~~100\%  &~48.1/43~~~27.4\%    \\
\hline
\end{tabular*}
\end{table}
 
The \babar~data~\cite{babar2013} from 2013 extend up to 5 GeV. We
have used only 48 energies up to 1.065~GeV, which is the highest
energy in the CMD-3~\cite{cmd2018} data. The \babar~data also cover the 
low-energy region and provide 13 data points between the $\kpkm$ 
threshold and the lowest CMD-3~\cite{cmd2018} energy. They are therefore
more sensitive to the subthreshold behavior [given mainly by the 
position(s) and residuum (residua) of the pole(s)] of the reaction amplitude.
Thanks to that, we have gotten the following results (see the rightmost panel
of Table~\ref{tab:twoexperiments}): the ``no kaonium'' and ``Coulombic
kaonium'' hypotheses are rejected, and the kaonium with a binding energy
of 10~MeV gives an acceptable fit. The details of the fits are given in
Table~\ref{tab:babar}.
\begin{table}[t]
\caption{\label{tab:babar}Parameters of the fits to the $\sqrt
s<1.065$~GeV subset of \babar~$K^+K^-$ data~\cite{babar2013} assuming 
no pole below the threshold (second column) and the 2p kaonium  with the 
Coulombic binding energy $b$ (third column) or with $b=10$~ MeV (fourth column).}
\begin{tabular*}{8.6cm}[t]{lccc}
\hline
                &~~~~No kaonium~~~~&~$b=1.64$~keV~&~~~$b=10$~MeV~~~\\
\hline
$R_1$ (GeV$^2$) & 0.3577(17)    & 0.3575(17) & 0.3572(18)   \\
$M_1$ (GeV)     & 1.019144(17)  & 1.019120(17) & 1.019034(19) \\
$\Gamma_1$ (MeV)& 4.461(37)     & 4.455(37)  & 4.300(40)    \\
$R_2$ (GeV$^2$) &              & 0.0140(13) & 0.184(18)    \\
$M_2$           &               & $2m_{K^+}-b$ & $2m_{K^+}-b$ \\
$\Gamma_2$      &               & 0 (fixed) & 0 (fixed)           \\
$\delta_2$      &               & --0.07(18)  & --1.070(56)    \\
\hline
\end{tabular*}
\end{table}

\section{2$\bm{p}$ kaonium in the $\bm{\die\to\pippim}$ data}
\begin{figure}[t]
\includegraphics[width=8.6cm]{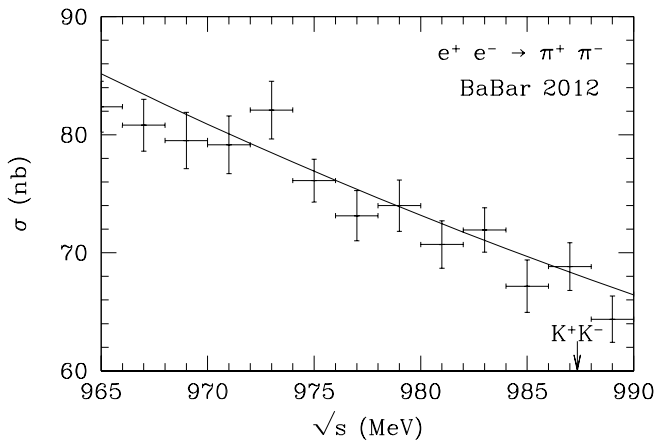}
\caption{\label{fig:bab12kaonium}Cross section for the $\die$ annihilation
into two charged pions measured in the \babar~experiment~\cite{babar2012}.
The curve is our fit to data from 0.3 to~1.6 GeV ($\cndf= 257.6/303$,
C.L.~$=97.2$\%). The $K^+K^-$ threshold is marked with an arrow.}
\end{figure}

There is also another possible way of seeing the 2p kaonium.
Thanks to its $\pi^+\pi^-$ decay mode, the 2p kaonium may, in principle, 
reveal itself as an irregularity in the $\die\ra\pippim$ excitation function 
slightly below $\sqrt s=2m_{K^+}$. The contemporary experimental data 
do not show any anomaly that could be interpreted as a manifestation of
2p kaonium. The invariant energy region in question has been ignored by most 
experiments, which have concentrated on the measurement in the 
$\rho/\omega$ region or at energies higher than 1~GeV. The important
exception is the \babar~experiment~\cite{babar2012}, which covered energies 
ranging from 0.3 to 3 GeV by using the initial state radiation 
method. Their data in the presumed signal region 
are shown in Fig. \ref{fig:bab12kaonium}, together with our ``conservative''
fit that does not include kaonium. There is an insignificant excess at 
$\sqrt s\approx 973$~MeV, but no firm conclusion can be drawn yet. More precise 
and denser data would be necessary to confirm or reject the presence of 
2p kaonium in the $\die\to\pippim$ data. According to
\cite{cmd3prelim}, we may expect new data from the CMD-3 experiment soon.
Figure~4 in Ref.~\cite{cmd3prelim} hints at their covering energies
to almost 1 GeV.

\section{2$\bm{p}~K^0$-onium as a pole in the $\bm{\die\to K^0_SK^0_L}$ 
amplitude}

The result we obtained when exploring the $\kpkm$ \babar~data \cite{babar2013} 
indicates that the strong interaction between a kaon and its antiparticle 
is attractive. One 
may therefore speculate about the existence of the $K^0\bar{K}^0$ bound state.
This would be the first (and probably only) onium composed of neutral
particles. To pursue this idea, we use the high-precision data on the
$\die\ra K^0_SK^0_L$ process~\cite{cmd2016}, again coming from the CMD-3
experiment at Budker Institute of Nuclear Physics in Novosibirsk, Russia, and 
comprising of 25 data points in a narrow interval around the 
$\phi(1020)$ resonance. 

Our simple one-resonance fit (the middle column in Table \ref{tab:parameters0}) 
provides $\cndf=36.7/22$, C.L. = 2.5\% \cite{notecmd2016}. The sophisticated 
fit by experimentalists~\cite{cmd2016} is much better ($\cndf=15/21$, 
C.L. = 82\%).
To explore the one-resonance--one-bound state scenario, we choose again
the Uretsky-Palfrey value of 10~MeV for the binding energy $b$. 
The result of the fit is $\cndf=7.5/20$, C.L. = 99.5\%, and
all parameters are listed in the rightmost column 
in Table~\ref{tab:parameters0}. Keeping in mind a good fit by the
experimenters themselves, we can say that the $K^0$-onium is not required 
by the CMD-3 \cite{cmd2016} data. When looking for $K^0$-onium in the 
$\pi^+\pi^-$ data, the signal region should be shifted upward by about 
7.9 MeV.
\begin{table}[t]
\caption{\label{tab:parameters0}Parameters of the two fits to the CMD-3 
$K^0_S K^0_L$ data~\cite{cmd2016}.}
\begin{tabular*}{8.6cm}[t]{lcc}
\hline
                &~~~~~~~~~One resonance~~~~~~~~~&Resonance and pole\\
\hline
$R_1$ (GeV$^2$) & 0.3488(14)    & 0.3500(39)   \\
$M_1$ (GeV)     & 1.019221(14)  & 1.019263(16) \\
$\Gamma_1$ (MeV)& 4.144(28)     & 4.168(29)    \\
$\delta$        & 0 (fixed)     & 3.2$\pm$2.0  \\
$R_2$ (GeV$^2$) & 0 (fixed)     & 0.0289(53)  \\
$b$ (MeV)       &     & 10.0\footnote{$M_2=2m_{K^0}-b$} (fixed)\\
$\Gamma_2$      &               & 0 (fixed)    \\
$\chi^2$        & 36.7          & 7.5         \\
NDF             & 22            & 20           \\
Conf. level     & 2.5\%         & 99.5\%       \\
\hline
\end{tabular*}
\end{table}

\vspace{11pt}
\section{Conclusions}
To conclude, we have found an indication of the existence of  kaonium
made of charged kaons in the 2p state by analyzing the data on the
$\die\ra\kpkm$ process. A similar analysis of the \nreact~data has provided
somewhat weaker evidence for the 2p $K^0$-onium. To this point, we have 
ignored the fact that both 2p kaonia have the same quantum
numbers. As a consequence, the $K^+K^-$ kaonium should be present in the
\nreact~amplitude as a bound-state pole, and the kaonium composed of neutral
kaons should reveal itself in the \chreact~amplitude as a bound-state pole 
or a resonance, depending on its bounding energy. Both 2p kaonia decay into
two charged pions. To get a clear and consistent picture of kaonia, the 
information should be combined from the $\die$ annihilation data to the 
$K^+K^-$, $K_SK_L$, and $\pi^+\pi^-$ final states. Work in this
direction is in progress.

\begin{acknowledgments}
I thank Filip Blaschke, Martin Blaschke, and Josef Jur\'{a}\v{n} 
for the useful discussions.

This work was partly supported by  Ministry of Education, Youth and
Sports of the Czech Republic Inter-Excellence Projects No. LTI17018 and
No. LTT17018.
\end{acknowledgments}

\end{document}